\documentclass[reprint,superscriptaddress,showkeys,amsmath,amssymb,aps,twocolumn,longbibliography]{revtex4-2}

\usepackage{graphicx} 
\graphicspath{{Figs/}}

\usepackage{bm} 
\usepackage{upgreek} 

\usepackage{fourier} 

\usepackage{graphicx}
\usepackage{multirow}
\usepackage{amsmath,amssymb,amsfonts}
\usepackage{amsthm}
\usepackage{mathrsfs}
\usepackage[title]{appendix}
\usepackage{textcomp}%

\usepackage[separate-uncertainty=true, multi-part-units=single]{siunitx}

\usepackage[version=4]{mhchem}  

\usepackage{xcolor} 
\usepackage{hyperref}
\hypersetup{colorlinks,breaklinks,
	urlcolor=blue,
	linkcolor=blue,
	citecolor=blue}

\usepackage{soul} 

\usepackage{fancyhdr} 
\usepackage{titlesec}
\titleformat{\section}[runin]{\normalfont\bfseries\itshape}{\thesection}{0.5em}{}[\,\,\,--]
\titlespacing*{\section}{0pt}{*1}{5pt}

\begin{document}
\title{Decoupling Entropic and Enthalpic Elasticity in Crosslinked Polymer Nanocomposites}
\author{
\centering\normalsize
Chiranjit Majhi,
Sivasurender Chandran,
and Krishnacharya Khare$^{*}$\\
\vspace{5pt}\small
\textit{Department of Physics, Indian Institute of Technology Kanpur, Kanpur - 208016, India}\\
\vspace{5pt}\normalsize
\textit{Email:}
$^*$\href{mailto:kcharya@iitk.ac.in}{kcharya@iitk.ac.in}
\vspace{5pt}
}

\begin{abstract} \bf
	
	Nanoparticle-induced local interface regions significantly enhance the bulk mechanical behavior of polymer nanocomposites, yet their energetic contribution in crosslinked elastomer networks remains poorly understood. Here, we decouple the entropic and enthalpic contributions to the elasticity of crosslinked poly(dimethylsiloxane) nanocomposites using temperature-dependent thermomechanical measurements. While the measured bulk elastic modulus exhibits a monotonic increase with nanoparticle volume fraction $\phi_{\rm n}$, we observe that the effective crosslink density initially increases before saturating at larger $\phi_{\rm n}$. These results are independently validated across two nanoparticle systems (silica and titania) and directly confirmed by X-ray scattering experiments on silica-based composites, highlighting the generality of the observation. The trends in effective crosslink density indicate that the elastic response is predominantly entropy-driven at low nanoparticle concentrations, whereas enthalpic contributions become increasingly significant at higher loadings. Assuming a linear superposition of interface and bulk contributions, these results reveal that the effective crosslink density at the particle interface decreases by an order of magnitude below a critical mean interparticle separation, indicating that localized particle proximity and clustering constrain the entropic elasticity at higher loadings.

\end{abstract}

\maketitle

\appendix
\begin{center}
\textbf{\Large \noindent Introduction}
\end{center}

Polymer nanocomposites have emerged as a versatile class of advanced materials because of their enhanced mechanical stiffness, toughness, and multifunctional behavior \cite{ahn2018multifunctional, swain2025harnessing}. These macroscopic reinforcements are fundamentally governed by the interfacial interactions between nanoparticles and the surrounding polymer chains \cite{kumar2013nanocomposites, chandran2014dispersion}. Chains adjacent to the nanoparticle surface experience strong physical or chemical pinning, resulting in altered conformations, restricted segmental mobility, and modified relaxation dynamics relative to the bulk matrix \cite{sen2007chain, sarkar2025nanotetrapods}. This perturbed interfacial zone – commonly treated as a distinct “interphase” – exhibits a gradual transition in the physical properties bridging the nanoparticle surface and the unperturbed polymer matrix \cite{qiao2011effect, fernandez2021immobilized}. 
Over the past two decades, extensive experimental and theoretical efforts \cite{fukahori2005new, wang2022polymer, huang2022interphase}, including atomic force microscopy \cite{huang2017local, zhang2018determination}, dielectric spectroscopy \cite{ghanbari2012interphase}, nuclear magnetic resonance \cite{krupanandam2026understanding}, and X-ray scattering \cite{holt2014dynamics, sharma2015structure} have been devoted to understand the spatial extent and local dynamics of this interphase. However, the vast majority of these investigations have focused on noncrosslinked polymer melts or solutions, where the rheological behavior is predominantly governed by transient entanglements and cooperative segmental flow \cite{bansal2005quantitative, srivastava2007experimental, jouault2013bound, ondreas2019effect, emamy2020structural}. In contrast, permanently crosslinked polymer networks remain comparatively underexplored, despite their critical technological importance in several applications, including structural damping\cite{chen2022research,li2025reversibly} and flexible electronics \cite{li2015flexible, li2023review}.

In crosslinked polymer systems, a permanent topological architecture imposes strict entropic constraints on chain conformations, dictating both local stress transmission and macroscopic rubber elasticity \cite{flory1960elasticity, shen2018effects, chen2020nanoparticle, ain2024role}. Introducing nanoparticles into such networks introduces a complex convolution of several effects: i) localized stress concentrations \cite{atif2016degradation, young2016interfacial}, ii) perturbation of crosslink networks \cite{lin2010development, kubiak2019forming}, and iii) immobilized physical regions mimicking permanent topological crosslinks/constraints \cite{suresh2019critical, chen2020nanoparticle, alasfar2022review}. Therefore, while macroscopic mechanical characteristics – stiffness and modulus – are known to increase monotonically with nanoparticle loading, conventional mechanical measurements alone do not distinguish whether this reinforcement is driven by an increase in effective entropic crosslink density or by direct enthalpic interactions (filler–polymer or filler–filler).  

Addressing this challenge requires isolating the entropic and enthalpic components of network elasticity over a wide range of particle loadings. Here, using temperature-dependent thermomechanical measurements, we directly decouple the entropic and enthalpic contributions to the elastic modulus of crosslinked polydimethylsiloxane (PDMS) nanocomposites filled with PDMS-grafted silica and titania nanoparticles. While the elastic modulus increases systematically with the nanoparticle volume fraction ($\phi_{\rm n}$), the effective crosslink density increases initially before saturating at higher concentrations, directly reflecting a concurrent reduction and eventual plateauing of the mesh size with $\phi_{\rm n}$. This mesh-size evolution is independently validated by X-ray scattering experiments on silica-based nanocomposites, confirming the decoupling of enthalpic and entropic components. An effective-medium model, based on a linear superposition of crosslink densities within the unperturbed bulk and the interphase regions, reveals that the effective crosslink density at the nanoparticle interface decreases by an order of magnitude below a critical mean interparticle separation. These results demonstrate that localized particle proximity and clustering constrain the entropic elasticity at higher loadings, providing a quantitative framework for directly correlating nanoscale interfacial confinement with the macroscopic mechanical response of these elastomers.

\vspace*{0.5cm}

\appendix
\begin{center}
\textbf{\Large \noindent Experimental Section}
\end{center}

\textbf{X-ray scattering measurements:} 
Sylgard$\mathrm{^{TM}}$ 184 (Dow Chemicals, USA) was used as matrix for making crosslinked polymer network \cite{murphy2020tailoring}. Silica ($\mathrm{SiO_2}$, diameter 20 $\pm\;2$ nm; NanoAmor Inc., USA) and titania ($\mathrm{TiO_2}$, diameter 20 $\pm\;2$ nm; Merck, India) nanoparticles were used as rigid nanofillers. High-purity toluene (99.9\%, reagent-grade; Merck, India) was used as the solvent to prepare uniform PDMS-nanoparticle dispersions. 

\vspace*{0.4cm}
\textbf{Sample preparation:} PDMS-grafted nanoparticles were prepared via a one-step process, following established procedures \cite{chen2020one, celik2023sustainable}. Grafting densities of as-prepared particles, determined via thermogravimetric analysis using known core dimensions, are approximately 0.015 $\mathrm{chains/{nm}^2}$ for $\mathrm{TiO_2}$ particles and 0.018 $\mathrm{chains/{nm}^2}$ for $\mathrm{SiO_2}$ particles. Thin sheets (thickness $\approx 1$ mm) of crosslinked PDMS and PDMS containing $\mathrm{SiO_2}$ or $\mathrm{TiO_2}$ nanoparticles were prepared using a solution-casting protocol adapted from our previous work \cite{majhi2026engineering}. In summary, suspensions of as-prepared grafted nanoparticles and bare PDMS in toluene were stirred and sonicated to ensure homogenous dispersion. This mixture was subsequently heated to 70 $^\circ$C for 2 h to remove the solvent. Subsequently, the curing agent was added to the mixture at a 10:1 (w/w) base to crosslinker ratio. This viscous mixture was then degassed in vacuum to remove trapped bubbles before casting between glass plates separated by 1 mm spacers and cured at 75 $^\circ$C for 2 h. Homogeneous dispersion of nanoparticles within the PDMS matrix was confirmed through energy-dispersive X-ray spectroscopy mapping performed using a field-emission scanning electron microscope (JSM-7100F, JEOL, Japan) and reported in our previous study \cite{majhi2026engineering}. Both silica- and titania-based nanocomposite systems were prepared with nanoparticle loadings ranging from 0 to 15 wt\%. Weight fractions were converted to volume fractions using the mass densities of the cured PDMS ($ \approx 1.03~\mathrm{g/cm^3}$), amorphous silica ($ \approx 2.41~\mathrm{g/cm^3}$), and titania ($ \approx 4.26~\mathrm{g/cm^3}$).

\begin{figure*}[ht]
    \centering
    \includegraphics[width=1\textwidth]{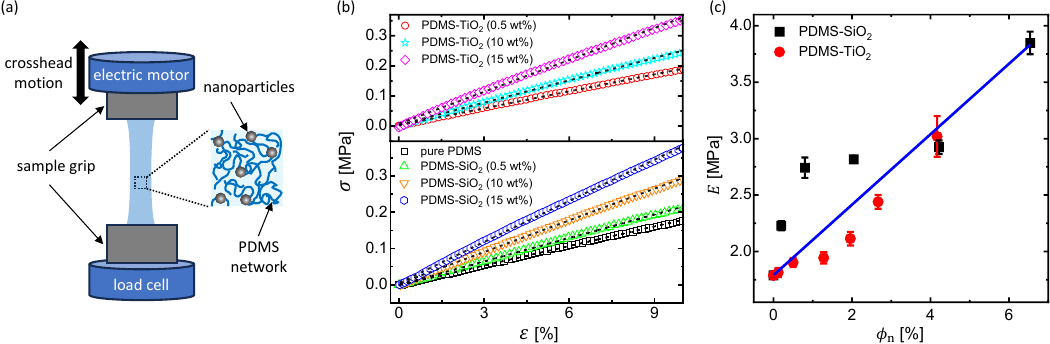}
    \caption {\textbf{Nanoparticle-induced mechanical reinforcement.} (a) Schematic illustration of the experimental setup for obtaining stress-strain response of nanocomposite sheets. (b) The stress-strain curves for pure PDMS and PDMS nanocomposites with the lowest (0.5 wt\%), intermediate (10 wt\%), and highest (15 wt\%) concentrations of $\mathrm{SiO_2}$ and $\mathrm{TiO_2}$ nanoparticles. These representative concentrations are selected to illustrate the mechanical response over the range. The elastic modulus is determined from the slope of the linear fits (black dashed lines) over the strain range up to $\varepsilon = 10\%$. The slope increases with nanoparticle concentration, indicating an enhancement in elastic modulus. (c) Bulk elastic modulus of PDMS nanocomposite films as a function of nanoparticle volume fraction for $\mathrm{SiO_2}$ and $\mathrm{TiO_2}$ nanoparticles. The error bars represent the standard deviation of the measurements from the set of five experiments, and the solid line represents the approximate linear dependence of $E$ on $\phi_{\rm n}$.}
    \label{fig-01}
\end{figure*}

\vspace*{0.4cm}
\textbf{Thermomechanical measurements:} Tensile stress-strain measurements were obtained using a universal testing machine (UTM; Z005, Zwick Roell, Germany), equipped with a precise load cell and temperature-controlled environmental chamber \cite{flory1949dependence, higuchi1948determination}. Rectangular specimens, 35 mm$\times$10 mm$\times$1 mm,  were gripped with an initial gauge length of 25 mm and stretched uniaxially at a cross-head speed 1 mm/min up to a maximum engineering strain of 20\% (See Fig.\ref{fig-01}a for the schematic). The Young’s modulus, $E$, was deduced from the slope of the resulting stress-strain curves in the linear response regimes ($\varepsilon=  10\%$) \cite{landau2012theory}. To investigate the bulk thermomechanical response and to decouple the thermodynamic contributions, measurements were performed over a range of temperatures: $T = 27-120~^\circ\mathrm{C}$. Before mechanical loading at each point, specimens were thermally equilibrated at the hold temperature for 10 min.

\vspace*{0.4cm}
\textbf{X-ray scattering measurements:} X-ray scattering measurements, in transmission mode, were performed on pristine PDMS and nanocomposites using SAXSpoint 5.0, (Anton Paar GmbH, Austria), operating with Cu-$K_\alpha$ source ($\lambda = 1.54~\text{\AA}$, photon energy $8.05$~keV) and a beam diameter of $100~\upmu\mathrm{m}$. Measurements were carried out at a sample-to-detector distance 500 mm, providing access to a range of scattering-vectors $q$: 0.04 to 6.5 $\mathrm{nm^{-1}}$. Two-dimensional scattering patterns were collected with an EIGER2 2M hybrid photon counting detector, with pixel size $75\times75~\upmu\mathrm{m^2}$. The acquired scattering patterns were corrected for sample transmittance and background scattering. Azimuthally averaged one-dimensional scattering intensity profiles, $I(q)$, were extracted using SAXSanalysis 4.70 software (Anton Paar GmbH). The resulting scattering intensities were normalized by the transmission factor and sample thickness to obtain the intensity per unit thickness.

\vspace*{0.5cm}

\appendix
\begin{center}
\textbf{\Large \noindent Results and Discussion}
\end{center}

Representative engineering stress ($\sigma$) – strain ($\varepsilon$) curves of pristine PDMS and nanocomposites with selected filler fractions are shown in Fig. \ref{fig-01}b, illustrating enhanced mechanical response for both the composite systems. The elastic modulus $E$, extracted from the linear small-strain region, is shown as a function of the nanoparticle volume fraction $\phi_{\rm n}$ in Fig. \ref{fig-01}c. Corroborating earlier results \cite{camenzind2010structure, cordoba2023pdms}, $E$ increases monotonically with increase in $\phi_{\rm n}$, revealing substantial mechanical reinforcement for both the composite systems. However, noticeable deviations from the linear trend can be observed at smaller fractions of the nanoparticles, indicating that $\phi_{\rm n}$ alone cannot capture the microscopic reasons underlying the enhancement in the elastic response. These deviations suggest filler-induced structural perturbations altering the balance between entropic and enthalpic contributions to the elastic free energy of deformation. 
\begin{figure*}[ht]
    \centering
    \includegraphics[width=1\textwidth]{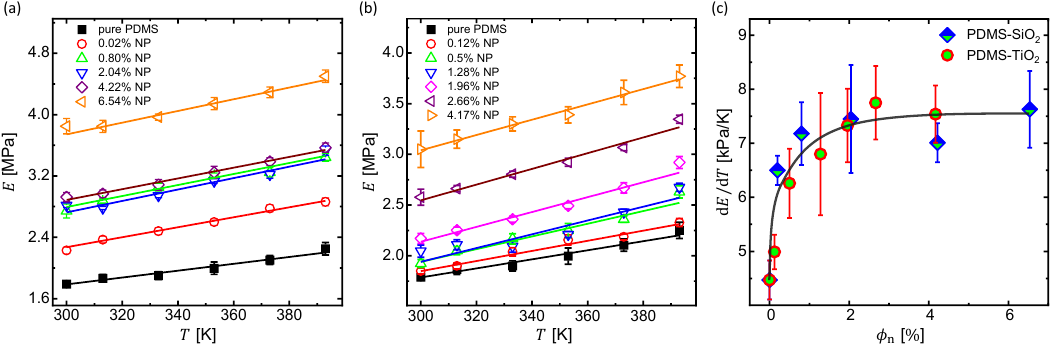}
    \caption {\textbf{Thermomechanical response of the nanocomposites.} Elastic modulus $E$ of pure PDMS and nanocomposites as a function of absolute temperature for (a)  PDMS-$\mathrm{SiO_2}$ and (b) PDMS-$\mathrm{TiO_2}$ nanocomposites. The error bars indicate the standard deviation from the set of five experiments. The linear fits to the data (solid lines) represent the temperature dependence of the elastic modulus, and the corresponding slopes quantify the sensitivity of the temperature-dependent elastic response ${\rm d}E/{\rm d}T$ for different nanoparticle concentrations in the PDMS matrix. (c) Variation in the slope of the temperature-dependent elastic modulus of PDMS nanocomposites as a function of $\phi_{\rm n}$. The error bars denote the standard error of fitting, while the non-linear solid black line is included as a guide to the eye to highlight the trend.}
    \label{Picture00TvsEs}
\end{figure*}
To decouple and quantify these distinct contributions, we investigate the thermomechanical behavior of the nanocomposites. Temperature-dependent $\sigma$ – $\varepsilon$ curves were obtained for both pristine and composite networks over a broad range of temperatures. Consistent with the classic signature of the entropic rubber elasticity \cite{wolf1975energetic, rubinstein2003polymer, mckenna2018soft, wang2020interfacial}, the elastic modulus increases linearly with temperature for all fractions (Fig. \ref{Picture00TvsEs}a \& b). To isolate the nanoparticle-induced effects on network thermomechanics, the temperature-dependence of elastic modulus $\mathrm{d}E/\mathrm{d}T$ is shown as a function of $\phi_{\rm n}$ (see Fig. \ref{Picture00TvsEs}c). For both types of composite systems, $\mathrm{d}E/\mathrm{d}T$ increases sharply at low $\phi_{\rm n}$ followed by a saturation at higher nanoparticle fractions. Because $\mathrm{d}E/\mathrm{d}T$ directly reflects the entropic restoring force of the network, this saturation at higher $\phi_{\rm n}$ provides direct evidence that the entropic contribution saturates even though the total modulus continues to increase (see Fig. \ref{fig-01}).

\begin{figure}
    \centering
    \includegraphics[width=\linewidth]{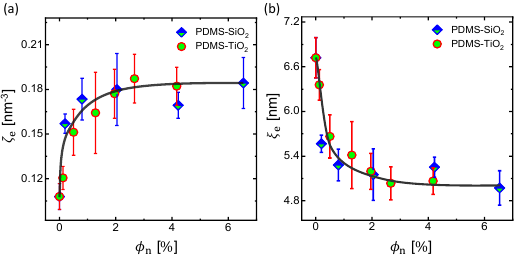}
    \caption {\textbf{Effective crosslink density and elastic correlation length of the composites.} (a) Effective crosslink density $\zeta_\mathrm{e}$ of the PDMS nanocomposite systems. Both systems exhibit a similar concentration-dependent increase at low nanoparticle loadings, followed by a saturation-like behavior at higher concentration $\phi_{\rm n}$. (b) Effective mesh size (elastic correlation length) $\xi_\mathrm{e}$ of the polymer network in PDMS nanocomposites as a function of $\phi_{\rm n}$. The error bars represent the standard deviations of the data. Both polymer-composite systems exhibit a similar concentration-dependent decrease at lower nanoparticle loadings, followed by a gradual approach to a saturation regime at higher concentrations. This indicates a common underlying balance between polymer-particle interfacial interactions and characteristic volume dilution effects. The non-linear solid lines are included as guides to the eye to emphasize the similar dependence of the both nanocomposite systems on $\phi_{\rm n}$.}
    \label{figEsTvsSf}
\end{figure}

To formalize this observation and to quantitatively decouple the entropic and enthalpic contributions, we examine the network free energy of crosslinked networks within the thermodynamic framework of rubber elasticity \cite{wolf1975energetic, rubinstein2003polymer}. The Helmholtz elastic free energy of the crosslinked nanocomposite network is expressed as
\begin{equation}
F_{\rm total}= F_{\text{entropic}}(\lambda,T,\phi_{\rm n},\epsilon_{\rm np}) + F_{\text{enthalpic}}(\lambda,T,\phi_{\rm n},\epsilon_{\rm np})
\end{equation}
where $\epsilon_{\rm np}$ characterizes the interaction between nanoparticles and the matrix polymer network, $\lambda=L/L_0$ is the extension ratio relative to the unstrained gauge length $L_0$. While the enthalpic contribution $F_{\text{enthalpic}} = U(\lambda,T,\phi_{\rm n},\epsilon_{\rm np})$ accounts for the localized bond distortions, filler--filler interactions, and energetic interactions at the nanoparticle--polymer surfaces, $F_{\text{entropic}}=-TS(\lambda,T,\phi_{\rm n},\epsilon_{\rm np})$ originates from the reduction in the conformational entropy of the elastically active polymer strands upon affine deformation. While the entropic contribution $F_{\mathrm{entropic}}$ scales linearly with the thermal energy ($k_{\rm B}T$), the enthalpic component varies negligibly for small deformations and modest temperature changes~\cite{mckenna2018soft}. Consequently, assuming the network is incompressible ($\Delta V/V \approx 0$), the tensile force on the network decomposes thermodynamically at constant volume and length as 
\begin{equation}
f = \left(\frac{\partial U}{\partial L}\right)_{T,V} - T \left(\frac{\partial S}{\partial L}\right)_{T,V} = f_{\rm U} + f_{\rm S}
\end{equation}
Applying the Maxwell relation $-\left(\frac{\partial S}{\partial L}\right)_{T,V} = \left(\frac{\partial f}{\partial T}\right)_{V,L}$ yields the classical thermodynamic equation of state for elastomer networks
\begin{equation}
f = f_{\rm U} + T \left(\frac{\partial f}{\partial T}\right)_{V,L}
\end{equation}
Expressing this relation in terms of the linear elastic modulus $E=\sigma/\varepsilon$, we obtain 
\begin{equation}
E = E_{\rm U} + E_{\rm S} = E_{\rm U} + T \left(\frac{dE}{dT}\right)
\end{equation}
According to the affine network theory, the entropic modulus $E_{\rm S}$ is governed by the number density of the elastically active strands, $ E_{\rm S} = 3\zeta_{\rm e}k_{\rm B}T$, providing the effective crosslink density as \cite{rubinstein2003polymer}
\begin{equation}
\zeta_{\rm e} = \frac{1}{3k_{\rm B}}\left(\frac{dE}{dT}\right)
\end{equation}
The effective crosslink density is shown as a function of $\phi_{\rm n}$ in Fig. \ref{figEsTvsSf}a. Strikingly, data from both silica- and titania-based nanocomposite networks collapse within the experimental uncertainty: $\zeta_{\rm e}$ increases sharply at low $\phi_{\rm n}$ before saturating at higher concentrations. 
To correlate these topological restrictions to structural length scales, we extract the number-average molar mass of the elastically active crosslinks as 
\begin{equation}
    M_{\mathrm{e}}=\frac{\rho_{\mathrm{pn}} \mathcal{N}_\mathrm{Av}} {\zeta_{\mathrm{e}}}
\end{equation}
where $\mathcal{N}_\mathrm{Av}$ is Avogadro's number, and $\rho_{\mathrm{pn}} = (1-\phi_{\mathrm{n}})\rho_{\mathrm{PDMS}}$ is the effective polymer matrix density. Modeling the sub-chains as Gaussian strands, the effective network mesh size $\xi_{\mathrm{e}}$, characterizing the structural correlation length between network constraints, is given by
\begin{equation}
    \xi_{\mathrm{e}} = l \left( C_n \frac{M_\mathrm{e}}{M_0} \right)^{1/2}
\end{equation}
where $l$ ($= 1.64~\text{\AA}$) is the segmental length, $C_n$ ($\approx 6.8$) is Flory's characteristic ratio for PDMS, and $M_0$ ($= 74.15~\mathrm{g/mol}$) is the molecular weight of the dimethylsiloxane repeat unit \cite{mahoney2002surface, rubinstein2003polymer}. The deduced effective mesh size decreases initially before flattening to a plateau at higher loading fractions (Fig. \ref{figEsTvsSf}b). Notably, data from both nanocomposite systems overlap, indicating that the results are governed by particle-induced conformational alterations at the interphase, rather than being nanoparticle-specific.


\begin{figure*}[ht]
    \centering
    \includegraphics[width=1\textwidth]{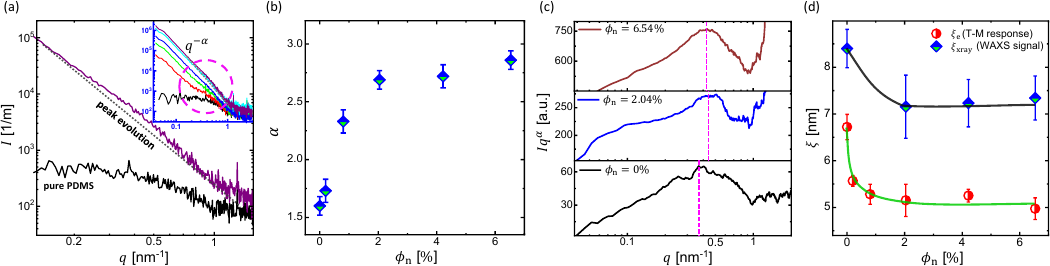}
    \caption {\textbf{Microscopic structural evolution via scattering.} (a) One-dimensional scattering profiles of pure PDMS and a representative PDMS-$\mathrm{SiO_2}$ sample. The power-law contribution ($I(q) \propto q^{-\alpha}$) is indicated by the dashed lines, where the scattering exponent $\alpha$ is determined independently for each sample. The deviation from this power-law behavior reveals the characteristic correlation peak. The inset shows one-dimensional scattering profiles of pure PDMS and PDMS-$\mathrm{SiO_2}$ nanocomposites with different nanoparticle volume fractions. (b) Variation of the power-law scattering exponent $\alpha$ as a function of nanoparticle volume fraction $\phi_{\rm n}$. (c) Scaled scattering intensity $Iq^{\alpha}$ as a function of $q$ for different samples, showing the characteristic peaks. The corresponding peak positions in reciprocal space are indicated by vertical dashed lines. (d) Characteristic correlation length $\xi_{\mathrm{xray}}$ as a function of $\phi_{\rm n}$. $\xi_{ \rm xray}$ is determined from the characteristic peak position $q^\ast$ of each sample. The correlation length obtained from X-ray scattering is compared with the characteristic mesh size $\xi_{\mathrm{e}}$ calculated from thermomechanical measurements. The error bars represent the standard deviation of the data, which is taken as the uncertainty of the measurement. The solid curves are included as guides to the eye to highlight the trends in the experimental data.}
    \label{figq0vsSf}
\end{figure*}

\begin{figure*}[ht]
    \centering
    \includegraphics[width=1\textwidth]{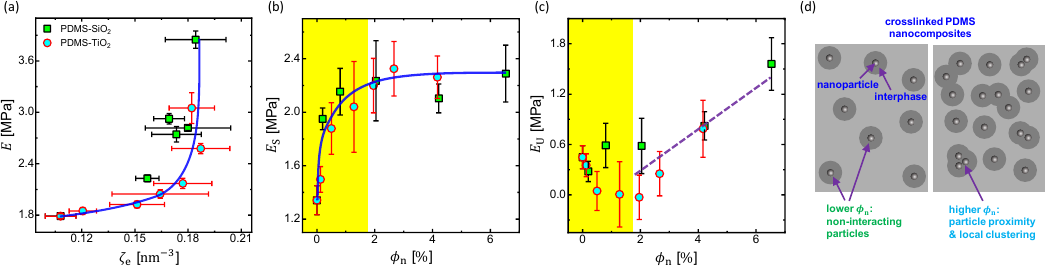}
    \caption {\textbf{Decoupling of entropic and enthalpic contributions to the total elastic modulus.} (a) Total elastic modulus at room temperature, $E$ is plotted as a function of effective crosslink density $\zeta_{\rm e}$, illustrating the concentration-dependent evolution of the elastic response. (b) The evolution of the entropic contribution $E_{\rm S}$ to the elastic response as a function of volume fraction $\phi_{\rm n}$. (c) Enthalpic contribution $E_{\rm U}$ to the total elastic response as a function of $\phi_{\rm n}$. The error bars represent the standard deviations of the calculated values. The solid curves in (a) and (b) are included as guides to the eye, whereas the dashed line in (c) represents the approximate linear dependence of $E_{\rm U}$ on $\phi_{\rm n}$ at higher concentrations. These results collectively demonstrate the concentration-dependent transition from an entropy-dominated elastic response to an enthalpy-dominated regime with increasing nanoparticle concentration. (d) Schematics representations of PDMS-nanocomposite illustrating non-interacting nanoparticles at low $\phi_{\rm n}$ and increased interparticle proximity and interfacial interactions at higher $\phi_{\rm n}$.}
    \label{ZevsEt}
\end{figure*}

To independently confirm this deduced evolution of mesh-network size and probe structural changes directly in reciprocal space, we turn to X-ray scattering measurements. Scattering intensity profiles, $I(q)$, for pristine PDMS elastomers and silica-based composite networks are shown in Fig. \ref{figq0vsSf}a. The pristine elastomer exhibits a transition from $q-$independent Guinier like plateau at low-$q$ to a power-law decay following $I(q) \sim q^{-\alpha}$, with $\alpha\approx 1.6 \pm 0.1$ characteristic of open, chain-like mass fractal network correlations (see Fig. \ref{figq0vsSf}b) \cite{feigin1987structure}. The crossover wavevector marking this transition reflects the intrinsic correlation length of the pristine network. Conversely, the composites reveal a power-law decay across the entire measured $q$ range, with $\alpha$ systematically evolving from $\approx 1.7$ to $2.9$ with increase in $\phi_{\rm n}$. This gradual evolution indicates a structural transition from open, polymer like fractals toward denser mass fractals. A similar transition in $\alpha$, saturating at $\approx$3.2, is observed in titania-based composites, underlining the generality of the structural transition. In addition to the power-law behavior, a subtle shoulder appears in the intermediate $ q$-range of the silica-based composites, signifying a characteristic structural correlation length. The weak amplitude of this feature stems from the relatively low electron density contrast between the crosslinked heterogeneities and the background PDMS matrix, in comparison to the dominant scattering from the inorganic nanoparticles. In titania-based composites, the much higher electron-density contrast of $\mathrm{TiO_2}$ cores causes the particle form-factor scattering to overwhelm this intermediate-$q$ region, masking the underlying polymer mesh feature. To isolate and extract these weak correlation peaks explicitly in silica-loaded nanoparticles, we construct Kratky-like compensated plots of $Iq^{\alpha(\phi_{\rm n})}$ vs $q$ in Fig. \ref{figq0vsSf}c. This presentation reveals a distinct correlation peak whose position, $q^\ast$ (Fig. \ref{figq0vsSf}c), shifts systematically towards higher $q$ with increase in $\phi_{\rm n}$. Converting the peak positions to the real-space correlation lengths $\xi=1/q^{\ast}$, we compare the scales derived from scattering with the mesh sizes deduced from thermomechanical measurements in Fig. \ref{figq0vsSf}d. The absolute scales extracted from scattering experiments are slightly larger as scattering captures total static spatial heterogeneities, whereas the thermomechanical measurements probe only the elastically active load-bearing strands. This agreement provides a direct, independent confirmation of the rapid reduction in network mesh size at low $\phi_{\rm n}$ and the saturation to a plateau at higher loading. This saturation of the effective crosslink density and the network mesh size at higher loading have an important mechanical consequence. In classic rubber elasticity, the entropic component of the elastic modulus scales linearly with the effective crosslink density, i.e., $E_{\rm s}\sim 3\zeta_{\rm e}k_{\rm B}T$. 
At low fractions, the measured modulus $E$ increases linearly with $\zeta_{\rm e}$ (see Fig. \ref{ZevsEt}a), confirming that the entropic elasticity governs the macroscopic mechanical response in this regime (highlighted by yellow shade in Fig. \ref{ZevsEt}b). The Enthalpic component to the elastic modulus $E_U$ remains concentration-independent in this region (also highlighted by yellow shade in Fig. \ref{ZevsEt}c). However, at higher nanoparticle loadings, $E$ continues to increase even after $\zeta_{\rm e}$ and the corresponding entropic modulus $E_S$ saturate. Consistent with this observation, $E_U$ exhibits an approximately linear dependence on $\phi_{\rm n}$ in this concentration regime (see Fig. \ref{ZevsEt}c). This trend reveals the growing dominance of enthalpic interactions driven by interparticle proximity and interfacial interactions (see Fig. \ref{ZevsEt}d). Together, these results demonstrate the $\phi_{\rm n}-$dependent crossover from entropy-dominated elasticity at low particle fractions to enthalpy-reinforced mechanical response at high loading.

\begin{figure}[ht]
    \centering
    \includegraphics[width=\linewidth]{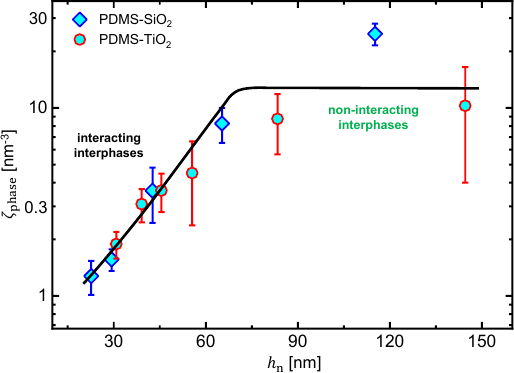}
    \caption {\textbf{Topological compaction and proximity-induced reductions in composites.} Both types of nanocomposite systems exhibit excellent overlap, as indicated by the black line, which serves only as a guide to the eye. The error bars represent the standard deviation of the data, which is taken as the measurement uncertainty. The trend in $\zeta_\mathrm{phase}$ shows two distinct behaviors: a plateau at larger $h_{\rm n}$ (non-overlap region) and a monotonic decrease at lower $h_{\rm n}$ (increasing-overlap region). These trends highlight the critical role of interphase overlap in governing the effective crosslink density of polymer nanocomposites.}
    \label{ZpvsSf}
\end{figure}

To rationalize the origin of this entropic saturation and isolate the topological response of the bound polymer layer, we construct an effective-medium model that partitions the composite into unperturbed bulk and particle-induced interphase regions. Assuming homogeneous dispersion, increasing nanoparticle volume fraction systematically decreases the mean surface-to-surface interparticle separation $h_\mathrm{n}$, following a geometrical relation \cite{srivastava2007experimental, chandran2014confinement},
\begin{equation}
h_{\mathrm{n}} = 2 r_{\mathrm{n}} \left[ \left(\frac{\phi_{\mathrm{rcp}}}{\phi_{\mathrm{n}}}\right)^{1/3} - 1 \right]
\label{eq08}
\end{equation}
where $r_{\rm n}\approx 10$ nm is the mean particle radius, and $\phi_\mathrm{rcp}=0.638$ represents the random close-packing volume fraction for monodisperse spheres in three dimensions. For all the fractions investigated here, $h_{\rm n}$ decreased monotonically from $\approx$~150 nm at the lowest particle fraction to $\approx$~20 nm at the highest loading. Considering the core diameter of $\approx$~20 nm and the characteristic hydrodynamic diameter of $\approx$ 8 nm for the grafted PDMS chains, this reduction brings neighboring grafted coronas into direct proximity at higher loading. 
Moreover, previous investigations on similar PDMS-$\mathrm{SiO_2}$ nanocomposites have demonstrated that the interphase regions extend over a characteristic size $d_{\mathrm{phase}}$ of approximately $4\text{--}5$ times the particle diameter, i.e., $d_{\mathrm{phase}}\approx 80-100$~nm \cite{huang2017local}. Consequently, as $h_{\mathrm{n}}$ decreases below this scale, interphases of the neighboring particles begin to overlap strongly, thereby diminishing the topological contrast between chains constrained near the particle interface and those in the unperturbed bulk.
To capture this behavior without assuming an arbitrary concentration-dependent shell thickness, we formulate a first-order level effective medium superposition in which the effective macroscopic crosslink density is partitioned between the crosslink densities near the interface ($\zeta_\mathrm{phase}$) and the unperturbed bulk matrix ($\zeta_\mathrm{bulk}$): 
\begin{equation}
    \zeta_\mathrm{e} = \phi_\mathrm{n} \zeta_\mathrm{phase}+ (1 - \phi_\mathrm{n}) \zeta_\mathrm{bulk}
    \label{eq09}
\end{equation}
Here, $\zeta_{\mathrm{bulk}}$ corresponds to the intrinsic crosslink density of pristine PDMS network ($\phi_{\mathrm{n}} = 0$), and $\zeta_{\mathrm{phase}}$ represents the effective crosslink density of the particle-perturbed domain (absorbing the local geometric interphase-to-core volume factor).

Using the bulk matrix value $\zeta_\mathrm{bulk}= 1.08 \times 10^{26}$ $\mathrm{m^{-3}}$ deduced from thermomechanical experiments on pristine polymers, we extract $\zeta_\mathrm{phase}$ via Eq.~\ref{eq09} for both composite systems and present it as a function of $h_{\rm n}(\phi_{\rm n})$ in Fig. \ref{ZpvsSf}. Across all fractions, the effective crosslink densities within the interphase are $10\text{--}100$ times higher than those of the unperturbed bulk matrix, reflecting significant topological compaction and chain immobilization near the nanoparticle surfaces. Notably, data from the $\mathrm{SiO_2}$ and $\mathrm{TiO_2}$ nanocomposites collapse onto a single master curve across the entire range of separations, revealing that the interfacial network architecture is governed by geometric confinement rather than the specific chemistry of the inorganic cores. At large separations ($d_{\rm n} + h_{\mathrm{n}} > d_{\mathrm{phase}}$, corresponding to dilute filler loadings), $\zeta_{\mathrm{phase}}$ exhibits a plateau, signifying isolated, non-interacting interphase shells surrounding individual dispersed nanoparticles. However, once the particle concentration increases such that  $d_{\rm n} + h_{\rm n} < d_{\rm phase}\approx100$~nm, $\zeta_{\mathrm{phase}}$ decreases exponentially by nearly an order of magnitude as $h_{\mathrm{n}}$ decreases further. This substantial drop indicates that overlapping interfacial zones between neighboring particles screen polymer chains, preventing them from forming independent entropic load-bearing bridges and effectively diluting the network density per particle. These results demonstrate that the entropic elasticity of the interphase is governed primarily by the degree of spatial overlap between neighboring interfacial domains.

\vspace*{0.5cm}

\appendix
\begin{center}
\textbf{\Large \noindent Conclusions}
\end{center}

Incorporating nanoparticles enhances the macroscopic mechanical response of crosslinked elastomers, yet the microscopic thermodynamic mechanisms governing this reinforcement have remained elusive. By performing systematic thermomechanical measurements across a range of temperatures on two different nanocomposite systems, we reveal a distinct decoupling of entropic and enthalpic contributions to the elastic modulus as a function of nanoparticle volume fraction $\phi_{\rm n}$. At low fractions, the rapid rise in effective crosslink density tracks the increase in the total measured modulus, confirming that the entropic network elasticity dominates the mechanical response. At high loadings, however, the elastic modulus continues to increase monotonically even after the saturation of the effective crosslink density, demonstrating $\phi_{\rm n}$-dependent crossover to enthalpic reinforcement. The saturation of crosslink density and the resulting decrease of the network mesh size towards a plateau at high loading is independently confirmed by X-ray scattering experiments. Moreover, the $q$-dependence of the scattering intensity captures a progressive structural evolution from open polymer-like correlations to compact mass-fractal-like architectures as $\phi_{\rm n}$ increases, underlining enhanced topological compaction within the composite systems. To link these structural changes to the macroscopic mechanics, we formulated an effective-medium theory that partitions the effective crosslink density into particle-induced interphase regions and unperturbed bulk-like regions. Our analysis within this framework suggests that isolated nanoparticles produce highly compact interphase regions with crosslink densities that are $\approx$~100 times larger than the unperturbed bulk matrix. Once the mean interparticle distance decreases below the characteristic interphase size ($d_{\rm n} + h_{\mathrm{n}} < d_{\mathrm{phase}}$), the interphase regions of the neighboring particles overlap. This, in turn, leads to an $\approx$10-fold decrease in the interphase crosslink densities at the highest probed $\phi_{\rm n}$. While this reduces entropic bridging, direct enthalpic interactions drive a continued increase in the elastic modulus. Overall, these findings and our quantitative effective-medium framework establish a clear microscopic link between nanoscale interfacial confinement and the macroscopic thermomechanical response of crosslinked polymer nanocomposites.
Considering a linear superposition of the crosslink densities of the bulk and interphase regions in determining $\zeta_{\rm e}$, the effective crosslink density of the particle interphase is found to vary monotonically with the interparticle spacing. This behavior suggests that decreasing interparticle spacing enhances overlap and interactions between neighboring interphases, leading to a reduction in the average crosslink density of the interphase regions. Overall, these findings provide insight into how nanoscale interfacial characteristics influence the macroscopic thermomechanical response of polymer nanocomposites.

\section*{\large References}

\bibliography{references.bib}

\end{document}